\documentstyle[preprint,aps,eqsecnum]{revtex}
\draft
\begin{document}
\title{Statistical properties of random scattering matrices}
\author{Petr \v Seba$^1$, Karol
\.Zyczkowski$^2$, and  Jakub Zakrzewski$^{2,3}$ }
\address{
$^1$ Nuclear Physics Institute, Czech Academy of Sciences, \\
250 68 Re{\v z} near Prague, Czech Republik }
\address{
$^2$Instytut Fizyki Mariana Smoluchowskiego, Uniwersytet  Jagiello\'nski,
\\ ul. Reymonta 4,  30-059 Krak\'ow, Poland}
\address{
$^3$Laboratoire Kastler--Brossel, Universit\'e
Pierre et Marie Curie,\\  T12, E1,
4 place Jussieu, 75272 Paris Cedex 05, France }
\date{\today}
\maketitle
\begin{abstract}
  We discuss the properties of eigenphases of $S$--matrices in
random models simulating classically chaotic scattering.
 The
energy dependence of the eigenphases is investigated and
the corresponding velocity and curvature distributions are obtained both
theoretically and numerically. A simple formula describing the velocity
distribution (and hence the distribution of the Wigner time delay)
is derived, which is capable to explain the algebraic tail of the time
delay distribution observed recently in microwave experiments.
A dependence of the eigenphases
on other external parameters is also discussed.
We show that in the semiclassical limit (large number of channels) the curvature
distribution of $S$--matrix eigenphases is the same as that corresponding
to the curvature distribution of the underlying Hamiltonian and is given
by the generalized Cauchy distribution.
\end{abstract}
\newpage
\narrowtext
\section{Introduction}

Quantum chaotic scattering has been discussed for a number of years
\cite{smil89,lewd91}. It may occur in a variety of different physical
situations from atoms and nuclei to  disordered mesoscopic devices
or microwave cavities. The schematic model of the system is presented
in Fig.\ref{scheme}. The cavity (internal region) is
coupled
to the outside world by leads. The characterictics of the internal
motion manifest themselves obviously in the properties of the $S$--matrix.
This is seen directly using the Hamiltonian approach to the scattering
\cite{mawd69}.

Consider a simple Hamiltonian with the Hilbert space spanned by $N$
discrete states $|k>$ and $M$ continua $|c,E>$:
\begin{equation}
\begin{array}{c}
H=\sum_{k=1}^{N} E_{k} |k><k| + \sum_{c=1}^{M} \int dE E |c,E><c,E| + \\
 \ \\
+ g\sum_{c=1}^{M}\sum_{k=1}^{N}\int dE \left (W_{kc}(E) |k><c,E| + h.c.
\right ).
\end{array}
\label{h1}
\end{equation}
Note that no continuum-continuum coupling is permitted in the model.
The bound-continuum coupling is characterized by coupling constant $g$
and the energy dependent matrix $W_{kc}(E)$, where it is assumed that
the columns of $W$ are normalized to unity.
If $W$ depends only weakly
on  the energy,
the corresponding unitary $M\times M$ matrix $S$ may be expressed as
\cite{mawd69,vwz85}
\begin{equation}
S_{cc'}(E)=\delta_{cc'}-2 g^2 i \sum_{kl}W_{ck}(E) \left (
\frac{1}{E-{\cal H}}\right )_{kl}W_{lc'}(E),
\label{smat}
\end{equation}
where ${\cal H}$ is the effective Hamiltonian describing the motion
within the bound subspace after eliminating the continua in a Markov
approximation. In an arbitrary basis spanning the bound subspace it takes
the form
\begin{equation}
{\cal H}_{kl} = H_{kl} - i g^2 \sum_{c=1}^M W_{kc} W_{cl}
\label{eff}
\end{equation}.

The question may be posed whether, for generic systems, there is a unique
relation between S-matrix properties and the type of motion inside
the cavity.
One way to address this issue is via the semiclassical theory \cite{smil89}
which seems, however, to be limited to a large number of channels, $M$.
On the other hand, in recent experiments on scattering in microstructures
\cite{mrwhg92}, \cite{cbpw94}
$M$ can be of the order of unity.

A second possible way is a stochastic approach in which the Hamiltonian $H$
and the coupling matrix $W$ \cite{lewd91} are modeled by random matrices.
A subsequent averaging over different realizations of $H$ (typically for fixed
$W$)  yields statistical
predictions concerning fluctuations of physical quantities of interest.
One hopes, extending the conjecture which has been quite useful for
bounded chaotic systems \cite{bohi91}
 that the properties of fluctuations
are universal.
For bounded, autonomous, classically chaotic systems,
 depending on their symmetries, the statistical
spectral fluctuations are, generically, well
represented by the corresponding quantities obtained from
 Gaussian Orthogonal (GOE), Unitary (GUE), or
Symplectic (GSE) Ensembles of random matrices \cite{mehta}.

Assuming that $H$ matrices are drawn from one of these ensembles and
making similar assumptions on the coupling matrix $W$ one is forced to ask
what are the properties of the ensemble of $S$--matrices. A partial
answer has been obtained by Lewenkopf and Weidenm\"uller \cite{lewd91}:
if $H$ belongs to GOE and the channels are equivalent (see below)
$S$ matrix belongs to COE provided the coupling constant $g$ equals to
unity.
This result has been recently generalized
to all three universality classes by Brouwer \cite{bro95} for arbitrary $g$.
 Brouwer found that the
$S$--matrices, Eq.(\ref{smat}), conform with Generalized Orthogonal
(GCOE) and Unitary (GCUE) Circular Ensembles
\cite{mps85} which
reduce to circular ensembles of Dyson (COE, CUE),
\cite{dys63} when the coupling to the continuum
becomes ideal. 

Brouwer's result provides
 a direct link with another popular, random matrix
theory based approach in which one draws directly $S$-matrices, or
transmission matrices, from an appropriate random matrix ensemble
\cite{mps85,imry86,mps87}. The Hamiltonian approach is, in
a sense, more general since it allows
for calculation of time delays and energy--correlation averages, while the ensemble of
$S$--matrices is energy independent. On the other hand, one is frequently
interested in single--point with respect to energy statistical
measures. Those may be directly accessed from the ensemble of $S$ matrices.
In such a way one may obtain, e.g., the universal conductance
fluctuations \cite{stone} from the random matrix model \cite{imry86}.

The study of mesoscopic devices points out that random
matrix approach has its limitations, e.g., the fluctuations may
become dependent on the length of the device \cite{jp95}. It is still,
an open question, what are the limits of the universality. We do
not adress this problem here, rather we want to study, within the
random model, the generic properties of $S$ matrices dependent on some
parameter. In this way we extend recent intensive studies of the
 statistical properties of bound systems dependent on the external
parameters \cite{gaspard,altshuler,zd93,vo94}.

 For parameter independent
Hamiltonians statistical properties are generically universal once
the mean level spacing, $\Delta$ is known. Parametric measures reveal a
similar universality, in addition to unfolding the energy levels
one has to unfold the parameter dependence \cite{gaspard,altshuler}.
On the other hand, much less is known about the parametric behaviour
of scattering systems. The two-point correlation function for
the $S$ matrix with respect to an external parameter was derived
recently by Mac\^edo \cite{am94}. The conductance fluctuations
in the presence of the magnetic field induced time-reversal
symmetry breaking have been studied by Pluha{\v r} et al. \cite{pwzl94}.
The semiclassical properties of the so called Wigner time delay have
been discussed by Jalabert and Pichard \cite{jp95}.
 Obviously, even
in the absence of the external parameter $S$ matrix is energy
dependent - the corresponding correlation function was obtained in
\cite{lewd91}.

Instead of investigating the $S$--matrix elements we focus on the
properties of the corresponding eigenphases (phase shifts).
Some important dynamical features of the system (for instance
the time delay inside the interaction region) can be easily expressed
using the derivatives of $S$--matrix eigenphases. Being directly
accesible in experiments, statistical properties of phaseshifts
deserve a detailed study. The nearest-neighbour spacing
distribution has been discussed in Ref.\cite{dsf91,ds92}. As
mentioned above we consider here parametric dependence of
phaseshifts.

The paper is organized as follows. The analytic results concerning
the distribution of velocities (i.e., first derivatives of
eigenphases with respect to energy or other external parameter), also
referred to as slopes, are discussed in Section II.
 Here we discuss also briefly the distribution
  of the second derivatives, i.e., the curvatures,
of the eigenphases.
 These predictions are tested against numerical results obtained
from simulations based on
Eqs.(\ref{h1}-\ref{smat}) (and their generalization allowing
for the presence of some external parameter)
in Section III. For systems dependent on the external parameter
one may also construct directly circular ensembles of
scattering matrices. The corresponding results are presented
in Section IV.

\section{Parametric Hammiltonian approach to $S$ matrix }
\label{s:param}

Consider the unitary matrix $S$ defined by equations (\ref{smat})
and (\ref{eff}).
To discuss properties of the eigenphases of the $S$ matrix
it is convenient to rewrite  Eq.(\ref{smat}) as
\begin{equation}
S(E)=\frac{1+iA}{1-iA},
\end{equation}
with $A$ given by
\begin{equation}
A=g^2 W^+\frac{1}{E-H}W.
\end{equation}

Eigenphases $s_m$ of the unitary matrix $S$
are related to  eigenvalues $a_m$ of the $M \times M$ Hermitian
matrix  $A$ by
\cite{bro95}
\begin{equation}
s_m=2{\rm arctan}(a_m)~~~~~ m=1,\dots,M.
\label{relat}
\end{equation}
Since $s_m$ is a function of $a_m$ the
statistical properties of $s_m$ and $a_m$ are identical after
unfolding.
 In the semiclassical limit (large $M$) it has been shown 
\cite{bro95} that $A$ belongs to the same ensemble as $H$.
 This implies
that the statistical properties (level spacing, number variance etc.)
of $s_m$ are identical with those of the eigenvalues of $H$.
Moreover, the relation (\ref{relat}) is useful when discussing the
parametric dependence of $s_m$.

To see how it works
consider the slopes of the eigenphases with respect to the energy $E$.
They  have a direct physical significance  as time delays associated
 with the corresponding phaseshifts
(see, e.g. \cite{jp95}) while the average slope is just the celebrated
Wigner time delay.

The eigenequation for $a_m$:
\begin{equation}
A|f_m>= g^2 W^+\frac{1}{E-H}W|f_m>= a_m|f_m>
\label{eq0}
\end{equation}
is equivalent to
\begin{equation}
\left(H+\frac{g^2}{a_m}WW^+\right)|h_m>=E|h_m>,
\label{eq}
\end{equation}
with $|f_m >$ and $|h_m>$ related by
$|h_m>=\frac{1}{E-H}W|f_m>$.
Note that the $N\times N$ eigenvalue problem Eq.(\ref{eq}) has, for
fixed $E$, only
$M$ nontrivial solutions $a_m$ and corresponding
eigenvectors $|h_m>$. This is related to the fact that $M-N$ eigenvalues
of $WW^+$ vanish. Differentiating Eq.(\ref{eq}) with respect to
$E$ for nontrivial $a_m$ one gets
\begin{equation}
\frac{g^2}{a_m^2}\frac{da_m}{dE}<h_m|WW^+|h_m>=<h_m|h_m>.
\label{norm}
\end{equation}
We shall assume from now on that $|h_m>$ are normalized to unity.

Using the relation (\ref{relat}) between $s_m$ and $a_m$,
 we express the inverse time delay as
\begin{equation}
u_m=\tau_m^{-1}=\frac{1}{ds_m/dE}=\frac{1+a_m^2}{da_m/dE}.
\end{equation}
Thus eliminating the derivative  via Eq.(\ref{norm}) we obtain
\begin{equation}
u_m= g^2 <h_m|WW^+|h_m>+g^2 \frac{<h_m|WW^+|h_m>}{a_m^2}.
\label{temp1}
\end{equation}
But it follows from Eq.(\ref{eq0}) and the relation between $|f_m>$ and
$|h_m>$ that the second term on the
right hand side above is proportional to the norm $<f_m|f_m>$.
Thus finally we get:
\begin{equation}
u_m= g^2 <h_m|WW^+|h_m>+<f_m|f_m>/g^2.
\label{inv}
\end{equation}
Eq.(\ref{inv}) indicates that the distribution of inverse time
delays is related to the norms appearing on its rhs.
Having in mind that $<h_m|WW^+|h_m>$ is a sum
of $M$ terms:
\begin{equation}
<h_m|WW^+|h_m>=|<h_m|w_1>|^2+ ... +
|<h_m|w_M>|^2,
\label{chiche}
\end{equation}
where $w_k,\ k=1..M$, are vectors describing the coupling
to the channel $k$, we can estimate the distribution of the
matrix element $<h_m|WW^+|h_m>$ by a $\chi^2$ distribution with $M$
degrees of freedom in the GOE case and $2M$ degrees of freedom in the GUE
case respectively. On the other hand $<f_m|f_m>$ has
a $\chi^2$ distribution with 1 (GOE case) or 2 (GUE) degrees of freedom.
Assuming the two terms on the r.h.s of Eq.(\ref{inv}) to be
independent, we obtain the distribution $P(u)$
by convolution of two $\chi^2_{\nu}$ distributions with different number
of degrees of freedom and different means. The result can be
expressed in terms of the confluent hypergeometric function $_1F_1(a;c;x)$,
sometimes called Kummer function \cite{atlas}.
Changing the variable to $\tau=1/u$ we obtain the distribution of time
delays

\begin{equation}
P\left(\tau\right)= { \exp [-  {g^2 \over \tau}]~
_1F_1 \bigl( {M \beta\over 2}; {(M+1)\beta \over 2};
 (g^2-g^{-2}){1\over\tau}\bigr) \over
  g^{\beta(M-1)} ~ \Gamma \left(\frac{\beta (M+1)}{2}\right)
\tau^{\beta\frac{M+1}{2}+1} },
\label{velkum}
\end{equation}
 where
$\beta=1,2$ for orthogonal and unitary ensembles, respectively.

Figure \ref{meangg} shows the mean time delay $\langle
\tau\rangle$ as a function of $g$  obtained by integration of the above
distribution for $\beta=2$ and $M=2,4,6$ and $16$. In general mean time
delay decreases with $M$. Moreover, the value of the
coupling constant $g_m$, for which
the mean delay time is maximal, decreases with the number of
channels. In both limiting cases
$g \to 0$ and
$ g \to \infty$ the mean time delay tends to zero, but the physical
meaning of this fact is different. In the former case the
coupling is so weak that the scattered wave is not affected by the
bound system $H$. In the latter case very strong
coupling causes the scattering to occur almost instantaneously.

Formula (\ref{velkum}) simplifies in the
 special case $g=1$. Then the inverse time delays are
distributed according to $\chi^2$ distribution
with $\beta(M+1)$ degrees of freedom and the mean $\langle u
\rangle=\beta(M+1)/2$.
The time delay distribution, $P(\tau)$ is then
\begin{equation}
P\left(\tau\right)=\frac{1}{\Gamma
\left(\frac{\beta (M+1)}{2}\right)}\tau^{-\beta\frac{M+1}{2}-1}
e^{-\frac{1}{\tau}},
\label{veldis}
\end{equation}
with the mean time delay
\begin{equation}
  \langle \tau \rangle=2/(\beta(M+1)-2).
\label{tauav}
\end{equation}

For large number of channels $M$ the $\chi^2$ distribution
can be approximated by a Gaussian with the variance equal to the mean
$\langle u \rangle$. Thus for $M$ large
\begin{equation}
\label{appr}
P(\tau)=\frac{2\alpha}{\sqrt{\pi
}\tau^2}\exp\{-\alpha^2(2/\tau-2/\tau_0)^2\},
\end{equation}
where
 $\alpha^2=1/(4\beta(M+1))$ and $\tau_0=2/\beta(M+1)$.

Consider now the case when
 the Hamiltonian $H=H(x)$ depends on
some external parameter $x$ and investigate the properties of $s_m(x)$
keeping the energy, $E$,  fixed.

Equation (\ref{eq}), putting explicitly the $x$-dependence,
takes the following form
\begin{equation}
\left(H(x)+\frac{g^2}{a_m(x)}WW^+\right)|h_m(x)>=E|h_m(x)>,
\label{eqx}
\end{equation}
and may be viewed as an eigenvalue equation for eigenvalues $E_n(a,x)$ of
 the matrix
$H(x)+\frac{g^2}{a}WW^+$,
 defined for any real parameter $a$.
Then Eq.(\ref{eqx}) is equivalent to a set of implicit
equations
\begin{equation}
E_n(a,x)=E\ ; \ \ \ n=1,2,..N
\label{eq2}
\end{equation}
for an unknown $a(x)$. Again
due to the positivity of $WW^+$ the equation (\ref{eq2}) has at most one
solution $a_m(x)$ for each $n$ with the total number of all possible
solutions being equal to $M$. Moreover from Eq.(\ref{eq2}) follows
\begin{equation}
\frac{da_m(x)}{dx}= \frac{\partial E_m(a,x)}{\partial
x}\frac{a_m^2}{g^2 <h_m|WW^+|h_m>},
\label{eq2a}
\end{equation}
where $E_m$ denotes those eigenvalues for which (\ref{eq2}) has
a nontrivial solution.
Combining (\ref{eq2a}) and (\ref{relat}) together with Eq.(\ref{norm})
yields simply
\begin{equation}
v_m=\frac{d}{dx}s_m =\tau_m\frac{\partial E_m}{\partial
x}
\label{eq3}
\end{equation}
providing the connection between the slope of the eigenphase with respect to
external parameter $x$, the corresponding time delay and the slope
of the {\it hermitian} eigenvalue problem Eq.(\ref{eqx}). The
generic distribution of slopes for Hermitian random matrices is
Gaussian-like \cite{altshuler,zd93}. Thus the eigenphases slope
distribution $P(v)$ may be obtained as a simple integral of the
Gaussian and $P(\tau)$ given by Eq.(\ref{veldis}) [we consider the critical
coupling, $g=1$ case only]:
\begin{equation}
P(v)=\frac{1}{\sqrt{2\pi}}\int_0^\infty P(\tau)
\exp\left(- \frac{v^2}{2\tau^2}\right)
\frac{d\tau}{\tau},
\label{P(v)}
\end{equation}
under the assumption that appropriate unfolding of the parameter, $x$,
has been made \cite{altshuler,zd93}. The integral in (\ref{P(v)})
yields \cite{rizik}
\begin{equation}
\label{pvvan}
P(v)=\frac{\beta(M+1)}{\sqrt{8\pi}v^{\beta(M+1)/2 +1}}
{\cal D}_{-\beta(M+1)/2-1}(1/v),
\end{equation}
where ${\cal D}_p(z)$ is a shorthand notation for the product of
the parabolic cylinder function, $D_p(z)$, \cite{rizik} with the
exponential
\begin{equation}
{\cal D}_p(z)=\exp(z^2/4)D_p(z).
\label{parab}
\end{equation}
Using the series representation of ${\cal D}_p(z)$ for small $z$ one
easily verifies that the large $|v|$ tail of $P(v)$ decays
algebraically as $P(v)\propto 1/|v|^{\beta(M+1)/2 +1}$. The
asymptotic form of ${\cal D}_p(z)=z^p[1+O(z^{-2})] $\cite{rizik}
 valid for large $z$ yields regular behaviour of $P(v)$
for $|v|$ small.

Let us recall that the distribution of level slopes with respect to an
external parameter, $x$, for a bounded generic system is Gaussian
\cite{altshuler,zd93}. It is remarkable that the corresponding slope
distibution for the eigenphases of the $S$ matrix reveals the
algebraic tails. As we shall demonstrate in the numerical example below,
for large $M$ the algebraic tail asymptotic form appears for
very large $|v|$ only, while the center of the distribution resembles
a Gaussian. One may verify, using the asymptotic form pf
 $P(\tau)$ given by
 Eq.(\ref{appr}) that the Gaussian distribution for $P(v)$ is
recovered
  in the $M\rightarrow\infty$ limit.

A similar analysis can be carried out also for curvature distribution.
In the limit of large $M$ it can be shown in such a way that the
curvature distribution tends to the generalized Cauchy one, as for the
bounded systems \cite{zd93}.

\section{Numerical results - Hamiltonian approach}

We discuss the numerical results obtained for the slopes
with respect to the energy, $E$, first. As discussed above they are
directly related to
possible time delays in the scattering process and are, therefore, of
particular interest.
The time delay distributions, $P(\tau)$, obtained numerically by
diagonalizing the $S$ (\ref{smat}) for various number of channels
$M$ and $g=1$ in the GUE case are presented
in  Fig.\ref{veleue} and compared with the  formula (\ref{veldis}).
The dimension of the internal Hamiltonian was taken
$N=100$,
energy $E$ was set to zero and the data where obtained from
$15000$ generated $S$ matrices.
Fine agreement of the numerical results with
the formula (\ref{veldis}) does not depend on the  energy and
the number of eigenstates, provided $N>>M$ and $E$ is not too large.
Figure (\ref{velgg}) shows a comparison of the
distribution $P(\tau)$ obtained for $g\ne 1$ and the formula
(\ref{velkum}). Again the agreement is quite satisfactory.

Analogous results for the GOE case and $g=1$ are
 plotted on the Fig.\ref{veleoe} (dashed line). The agreement
with
 the prediction (\ref{veldis}) is  not so
nice as in the GUE case. Especially for small $M$ a discrepancy is
significant.
Importantly, it may be removed when we use
the formula (\ref{veldis}) with {\it one additional degree of
freedom}, i.e.:
\begin{equation}
\tilde{P}\left(\tau\right)=\frac{1}{\Gamma\left(\frac{M+2}{2}\right)}
\tau^{-\frac{M}{2}-2}
e^{-\frac{1}{\tau}}.
\label{veldisgoe}
\end{equation}
The corresponding average time delay is given by Eq.(\ref{tauav})
with $\beta=1$ and $M\rightarrow M+1$ substitutions.
We are not able to give a plausible explanation for this additive degree
of freedom. The numerical results demonstrate clearly, however,
 that it is superior to Eq.(\ref{veldis})
and  leads to an
excelent agreement with the numerical data - see
 Fig.\ref{veleoe}.

The formula (\ref{veldisgoe}) implies that the tail of
the distribution for GOE behaves as $\tau^{-M/2-2}$. In particular,
for
$M=3$ we obtain for the time delay distribution
$P(\tau)\approx\tau^{-7/2}$, in a full agreement with the
experimental finding in \cite{alt95}.

Similar numerical tests may be also performed  for the velocity
distribution in the case of a parametric dependence in order to test the
formulae (\ref{P(v)}),(\ref{pvvan}).
  To this end we have to introduce a parametric dependence
into the basic model of Eq.(\ref{smat}-\ref{eff}). One can, in principle,
discuss several possible cases, the parameter $x$ may affect either
the bound system only [described by $H_{kl}$ in Eq.(\ref{eff})] or
the decay part, i.e. the bound-continuum coupling $W$, or both. In a
generic case, arguably, both parts of the effective Hamiltonian will be
affected. We consider the case when
the internal dynamics only is $x$ dependent, i.e., $H=H(x)$.
 This situation
corresponds directly to predictions obtained in the previous section.
A similar approach has been adopted in the treatment of time
reversal symmetry breaking influence on the conductance
fluctuations \cite{pwzl94}.

The parametric $x$-dependence  is taken in a generic form:
$H=H_1\cos(x)+H_2\sin(x)$ \cite{zd93} where $H_i$ are
drawn independently from the
{\it same} ensemble of random matrices.
 The trigonometric form assures
that the mean density of states remains the same for all $x$ and the
motion of bound levels as a function of $x$ is stationary \cite{zd93}.
The slopes have been calculated by a finite difference in $x$ from eigenphases
of $S$. As before more than 15000 $S$-matrices were used for the
averaging.

In the GUE case the agreement between the numerical data and the
analytic prediction for the slope distribution, Eq.(\ref{pvvan})
is again remarkable - see
Fig.~\ref{velxu}. Note strongly non-Gaussian character of the
obtained distribution. In the semi-logarithmic scale used in Fig.~\ref{velxu}
a Gaussian would take an inverted parabola shape.
The distributions show an algebraic tail
$P(v)\propto v^{-(M+2)}$ for small $M$. For $M$ large
a center of the distribution resembles a parabola (i.e., a Gaussian
distribution in the linear scale) followed by a straight
line in the semilogarithmic  plot indicating a regime of
exponential
behaviour, $P(v)\propto \exp(-\gamma |v|)$. The numerical data
are insufficient to detect a transition to a possible algebraic
tail for $|v|$ large. For $M$ even larger (not shown) the regime
of Gaussian behaviour broadens, the exponential behaviour moves to the tails
indicating the transiton to the semiclassical limit.

Consider now GOE internal dynamics.
Similarly as for the time delays -
the agreement with the numerical data is
admirable if the distribution (\ref{P(v)}) is evaluated with
the ``ad-hoc'' improved ${\tilde{P}}(\tau)$
given by Eq.(\ref{veldisgoe}) -- compare Fig.~\ref{velxo}.
As before  the ``proper''
${P}(\tau)$ [Eq.(\ref{veldis})] with $\beta=1$ used in arriving at
Eq.(\ref{pvvan}) leads to a theoretical prediction
with clear disagreement with the numerical data for small $M$
(not shown). The corrected expression is obtained from Eq.(\ref{pvvan})
by $\beta=1$ and $M\rightarrow M+1$ substitutions.
The distributions obtained are similar in shape
to those corresponding to GUE case, with the
 algebraic tail of the form $P(v)\propto v^{-M/2-2}$ seen clearly for $M$
 small. For the largest number of channels plotted $M=16$ we again
 observe the Gaussian center with exponential tails in similarity with
 the GUE distribution.

To complete the discussion of the parametric dependence in
the Hamiltonian based approach
let us pass now to the curvature distributions.
Numerical tests performed for the curvatures  $K=d^2s/dE^2$
are presented
in Fig.\ref{curve}. Note
the quite good agreement of the numerical data with the generalized Cauchy
distribution \cite{zd93}. The double-logarithmic scale used enhances
the small and large $|K|$ behaviour. For large $|K|$ the agreement is
excellent confirming the universality of the large curvature tail
behaviour also for the present scattering system. On the other
hand one observes a slight excess of small curvatures (and the
corresponding lack of "medium" curvatures). It is a clear indication
that $M=16$ is not sufficient to realize fully the semiclassical limit.
Using the analogy with the bound system level dynamics \cite{zd93}
one may conclude that avoided crossings between eigenphases
(as a function of the energy, $E$) are still partially isolated.

The curvature distribution with respect to the external parameter is
plotted in Fig.\ref{curvx}. Comparing to the previous case,
 a better agreement with
the generalized Cauchy distribution is observed indicating that
the semiclassical regime is reached faster when the motion of
eigenphases as a function of $x$ is considered. A similar
qualitative conclusion
may be reached considering
 the distribution of time delays and of the slopes
with respect to the parameter $x$.

 For $M$ smaller one observes stronger deviations from the
universal Cauchy distribution (not shown). The explicit
dependence on the number of open channels in that case has
obviously the same origin as the corresponding dependence in the
case of slope (or time delay) distributions.

\section{Random parametric $S$-matrix}

As mentioned in the introduction one often employs random ensembles
to directly model properties of $S$ matrices \cite{mps85,imry86,mps87}.
One may envision a similar approach for the study of the
parametric statistics. To this end one has to define
the parametric dependence of $S$ matrices directly,
 without reference to the underlying  bounded dynamics.
Obviously, there is some ambiguity here, there are several possible
choices. The ideal approach should be conceptually simple and,
at the same time, lead to the same distributions
as those obtained for the Hamiltonian based approach.
After all in both cases the
parameter independent matrices belong to generalized circular ensembles
\cite{bro95}. We shall see below that the simplest possibilities
agree with the Hamiltonian approach in the limit of large
number of channels (semiclassical limit) only.

The first approach proposed takes  $S$ matrices in the form:
\begin{equation}
S(x)=S_0 \exp( i x V)
\label{kar1}
\end{equation}
where $S_0$ is drawn from the appropriate circular ensemble (COE or CUE)
of a given rank $M$ while $V$ is a Hermitian matrix
(rank $M$),
independent of $S_0$ and drawn from GOE or GUE, respectively.
Then the eigenphases at $x=0$ are
the phase shifts of $S_0$. The   slopes and curvatures of $S$
are easily defined as the corresponding derivatives at $x=0$.

 Such a form of a parametric dependence is
"borrowed" from the typical form of a Floquet operator corresponding
to quantum maps (e.g., for a famous kicked rotator or for the kicked top
- for a detailed introduction to such problems see \cite{haake}).
While in \cite{saher} the form of the Floquet operator was
determined by the dynamics of the system, here we assume a random $S_0$,
as discussed above.

Eq.(\ref{kar1}) implies a Gaussian distribution of slopes, since
 the slopes of eigenphases
of $S$ at $x=0$ are given by diagonal elements of matrix $V$
represented in the eigenbasis of $S_0$.
The latter are Gaussian distributed as we have taken $S_0$
and $V$ independently. The Gaussian character of the velocity
distribution for the model (\ref{kar1}) is independent on the matrix
size $M$ and holds for both COE and CUE.
 Recalling the results obtained in the Hamiltonian model
we see that the statistical predictions for eigenphase slope
distribution coincide only in the  $M\rightarrow\infty$ limit.

The random $S$ matrices belonging to a given ensemble of random
unitary matrices (COE, CUE)
may be generated (for arbitrary $M$) by drawing the generalized
Euler angles from the appropriate probability distribution
\cite{kzmk94}. Similarly one may construct the way to introduce
the parametric dependence by considering the
infinitesimal changes of the generalized Euler angles
 \cite{kzmk95}. Our numerical results indicate that if one
assumes that {\it all} $M^2-1$ angles are affected by the perturbation
in a same way,
the distribution of slopes is {\sl Gaussian}  for arbitrary $M$.

Results received of $10^4$ random unitary
COE matrices with $M=40$
are presented in Fig. \ref{velkar}.
Gaussian character of the "velocity" distribution, obtained in this
case, is not at all typical of any parametric dynamics
defined for random unitary matrices. On contrary, one may
construct many kinds of parametric dynamics in the space of unitary
matrices leading to {\sl non-Gaussian} distributions $P(v)$.
Exemplary data, represented in the figure by triangles, where obtained
for a model in which only $M$ diagonal Euler angles have beed varied.
For small $M$
the distributions obtained are closer to
the results of
the Hamiltonian model (\ref{smat}) than the pure Gaussian distribution.
Similar results hold also for CUE case.

Let us consider now the second derivatives of eigenphases with
respect to $x$.
For a Floquet operator corresponding to the kicked top model,
 the parametric curvature distribution has been
studied already \cite{saher} and shown \cite{zd93} to obey
 the generalized Cauchy
distribution with quite a good accuracy
for all three universality classes. The numerical simulations
performed by us indicate that for $M$ large (semiclassical limit)
the same property holds for the $S$ matrix models described above.

The simple propositions discussed above failed to reproduce
the distribution of slopes with respect to the external parameter,
$x$, obtained for small $M$ within the Hamiltonian $S$-matrix model.
However this goal may be simply obtained by a slight modification
of assumption concerning the matrix $V$ in Eq.(\ref{kar1}). As
discussed above the distribution of slopes of $S$ matrix eigenphases
is equivalent to the distribution of diagonal elements of $V$ in
the eigenbasis of $S_0$. Instead of choosing $V$ to belong to
the appropriate for the given symmetry Gaussian random ensemble
one may chose $V$ from the ensemble defined by the desired distribution
of diagonal elements, Eq.(\ref{P(v)}). This defines the distribution
of $M$ elements of $V$ leaving the remaining $M(M-1)/2$ elements
undefined (in a chosen basis). This additional freedom may be
utilized to make this ensemble less artificial than it seems at first.
While this {\it aposteriori} procedure may be hardly described as
an elegant one, it gives an indication of the properties of the
required ensemble of $V$ matrices. In particular, following the
discussion presented in Section II, it is clear that the construction
of such an ensemble may involve scalar products of unnormalized
vectors in $M$--dimensional space [compare Eq.
(\ref{chiche})]. Hopefully
one may define the random ensemble for $S$ matrices of the form
(\ref{kar1}) which is universal, i.e., reproduces not only
the distribution of slopes but also of all other statistical
measures involving the parameter $x$. This would allow to
use the "direct" $S$ matrix approach instead of the Hamiltonian
one, also for the problems involving the external parametric
dependence.

\section{Concluding Remarks}

In this work we have considered the statistical properties of
derivatives of the $S$ matrix eigenphases with respect to the
energy, $E$, as well as with respect to some external parameter
$x$. Within the Hamiltonian approach we were able to derive,
using simple heuristic arguments, the analytic
distributions for the time delay  valid for arbitrary
number of open channels $M$ and arbitrary value
of the coupling constant $g$. An analytic distribution
of slopes with respect to an external parameter has been also
obtained.
 A comparison with the numerical
simulations have shown good agreement of the theory with
the numerical data in the CUE case. The discrepancies observed
for the COE case may be removed by an ''ad-hoc'' modification
of the proposed expression.

When the main part of this
work has been finished we have been informed
 that in the CUE case the same
formulae for the time delay distribution have been obtained by Fyodorov and
Sommers
 using supersymmetry calculus \cite{Fyodorov}. We hope
that the supersymmetric approach will be able in the future to
verify the proposed distributions also for the time-reversal
invariant (COE) model.

We have verified numerically that the curvature distribution
(i.e., the distribution of second derivatives of eigenphases
with respect to energy) obeys the generalized Cauchy distribution
\cite{zd93}
in the semiclassical limit ($M$ large).

Construction of the parametric dependence directly for the
random $S$ matrices belonging to a given random ensemble has
been shown to be ambiguous. The natural choices for the dynamics
lead to Gaussian velocity distribution for arbitrary number of
open channels. Then the agreement with the Hamiltonian based
approach is obtained in $M\rightarrow\infty$ limit only.

J.Z. thanks D.~Delande for discussions. We acknowledge interesting
exchanges with Y.~Fyodorov and are grateful for informing
us of his results prior to the publication.

Laboratoire Kastler Brossel, de l'Ecole Normale Superieure et de
l'Universite Pierre et Marie Curie, is unit\'e associ\'ee 18 du CNRS.
 This work was  supported by
the Polish Committee of
Scientific Research under grant
No.~2P03B~03810 (K.\.Z. and J.Z.) and by the GA AS No. 148409
(P.\v S.).


\begin{figure}
\caption{
Scheme of the scattering system.
 A system containing $N$ bounded states
and described
by a Hamiltonian $H$ is coupled via matrices $W$ and $W^\dagger$
to two waveguides with $M_1$ and $M_2$ open channels.
The scattering in the system is thus
characterized by a $M \times M$ matrix $S$, where $M=M_1+M_2$.
}
\label{scheme}
\end{figure}

\begin{figure}
\caption{Dependence of the mean time delay $\langle \tau \rangle$
on the coupling constant $g$ for $\beta=2$. Number of channels $M$
labels each curve.
}
\label{meangg}
\end{figure}

\begin{figure}
\caption{Time delay
distributions, $P(\tau)$, for GUE internal dynamics, $g=1$
and  for different number of open channels, $M$ as indicated in
the Figure. The numerical data (histograms) and the
theoretical distributions (solid lines), $P_2(\tau)$,
Eq.(\protect\ref{veldis}) are presented as a function of
$\tau/<\tau>$ to facilitate comparison of distribution shape
for different $M$. According to Eq.(\protect\ref{tauav}) $<\tau>$ is
inversely proportional to $M$.
}
\label{veleue}
\end{figure}

\begin{figure}
\caption{Time delay
distributions, $P(\tau)$, for GUE internal dynamics and $g.ne.1$:
a) $g=0.5,~M=4$; b) $g=2.0, ~M=6$.
Solid line represents the theoretical
prediction as given by Eq.(\protect\ref{velkum}).
   }
\label{velgg}
\end{figure}

\begin{figure}
\caption{  Time delay distribution,
$P(\tau)$, for time reversal invariant (GOE) internal dynamics,
$g=1$
and  for different number of open channels, $M$ as indicated in
the Figure. The numerical data (histograms) are compared with
both the analytical prediction,
  Eq.(\protect\ref{veldis}), (dashed lines) and the corrected
prediction,
  Eq.(\protect\ref{veldisgoe}) (represented by a solid line).
}
\label{veleoe}
\end{figure}

\begin{figure}
\caption{Numerical data for
distribution of slopes (with respect to external parameter) for
GUE internal dynamic and $g=1$ plotted in the semilogarithmic
scale. The theoretical prediction,
Eq. (\protect\ref{pvvan})
 is shown as a full line and the
number of channels is indicated in each graph. Note strongly
non-Gaussian shape of the distribution.
}
\label{velxu}
\end{figure}
\begin{figure}
\caption{Same as in the previous figure but
 for
GOE internal dynamic.
The theoretical prediction is obtained by the modification of
Eq. (\protect\ref{pvvan}) discussed in the text.
}
\label{velxo}
\end{figure}

\begin{figure}
\caption{Distribution of second derivatives of eigenphases of $S$
matrix with respect to energy,
 $K=d^2s/dE^2$ obtained numerically
for the case of  $M=16$ open channels, (a) for GUE internal dynamics,
(b) for GOE dynamics. The solid (dashed) lines in both panels represent
the generalized Cauchy distributions corresponding to the GOE (GUE) case
respectively.
}
\label{curve}
\end{figure}

\begin{figure}
\caption{As in the previous figure for the curvature
with respect to an external parameter,
 $K=d^2s/dx^2$.
}
\label{curvx}
\end{figure}

\begin{figure}
\caption{Distribution of slopes of eigenphases
with respect to the perturbation parameter $x$
of random unitary matrices of size $M=40$
typical of COE; a)
 parameter $x$ controls variations
  of all $M^{2} -1$ Euler angles $(\triangle)$;
b)  only $M$ diagonal angles are varied $(\diamond)$.
}
\label{velkar}
\end{figure}

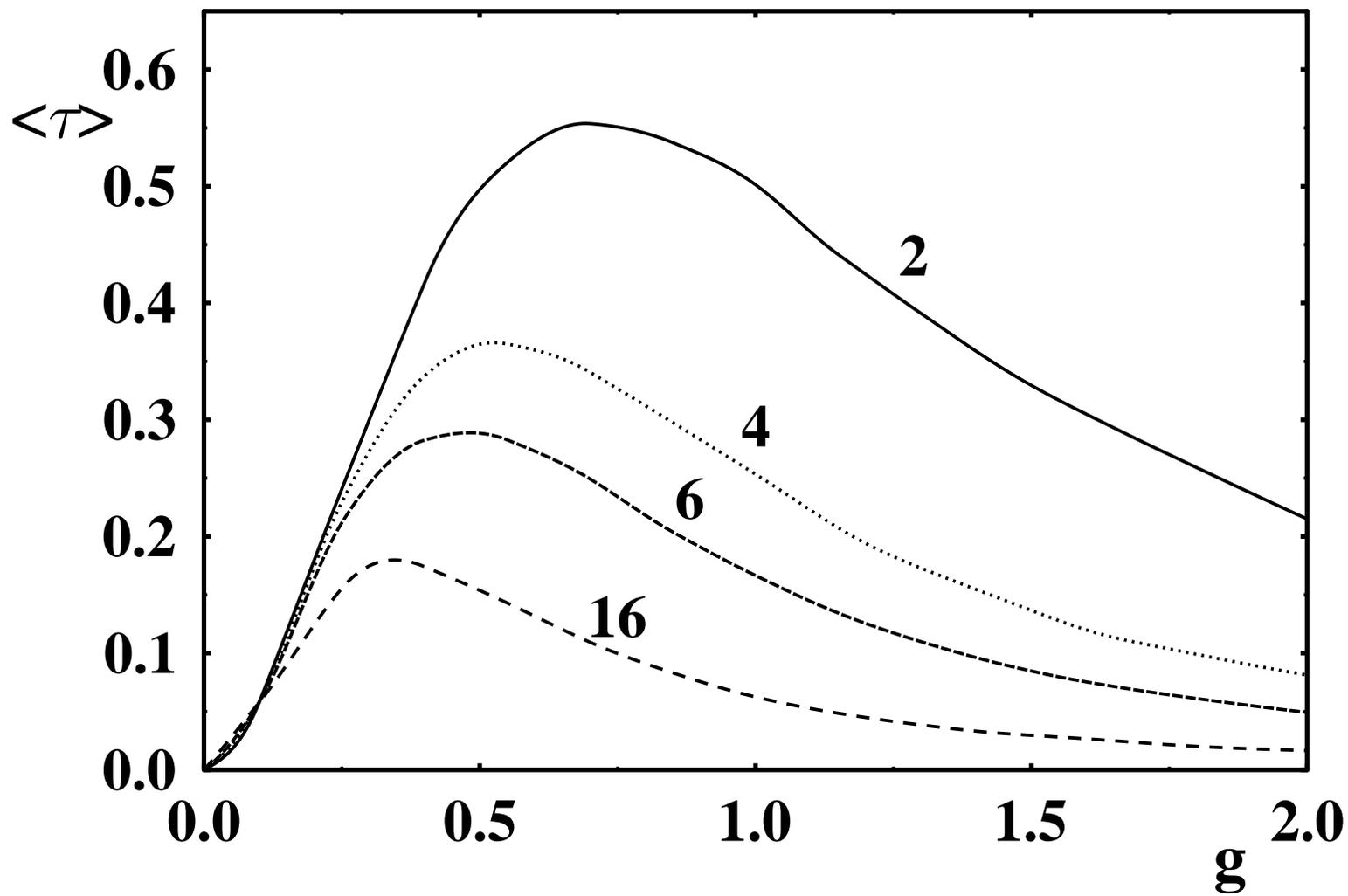
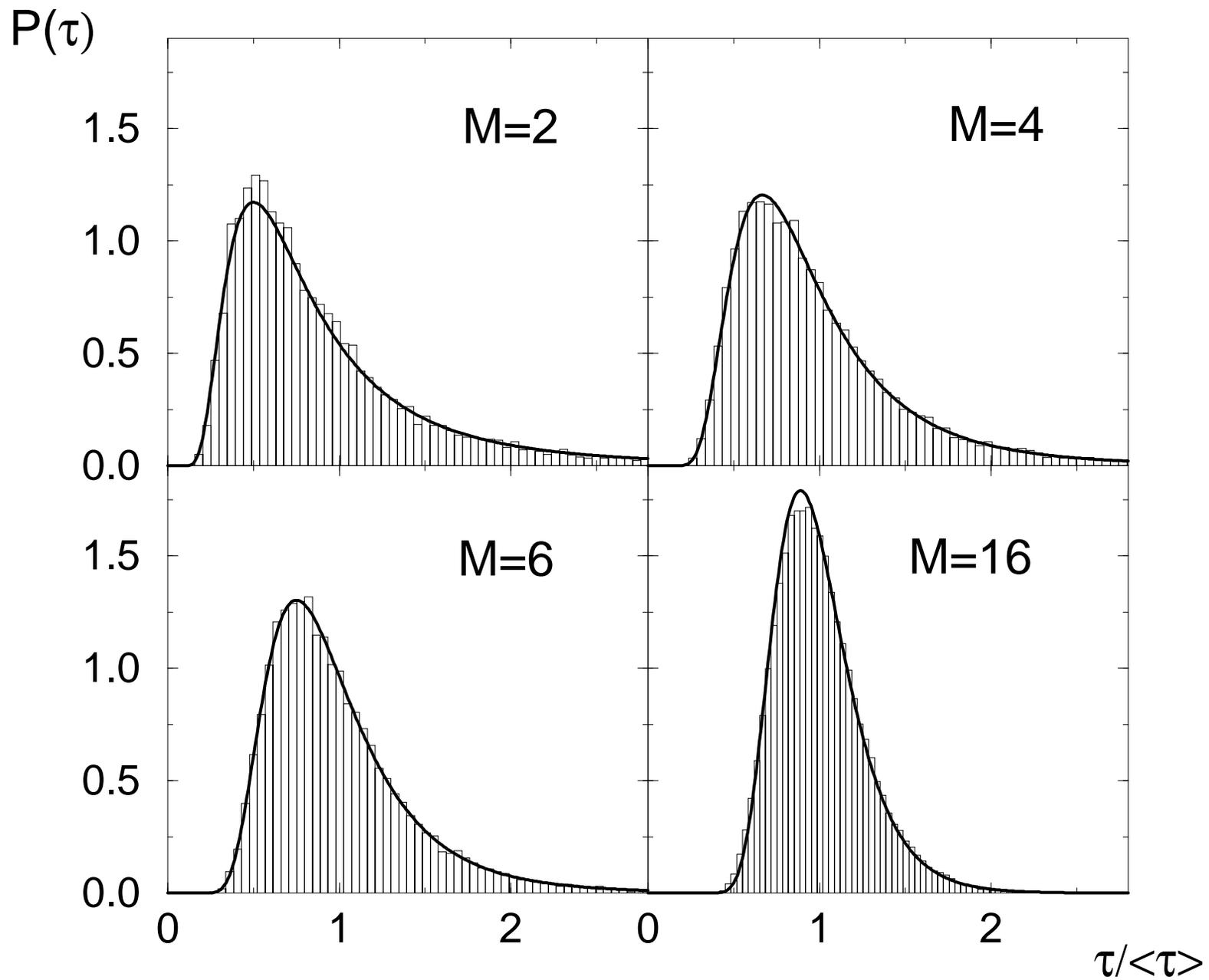
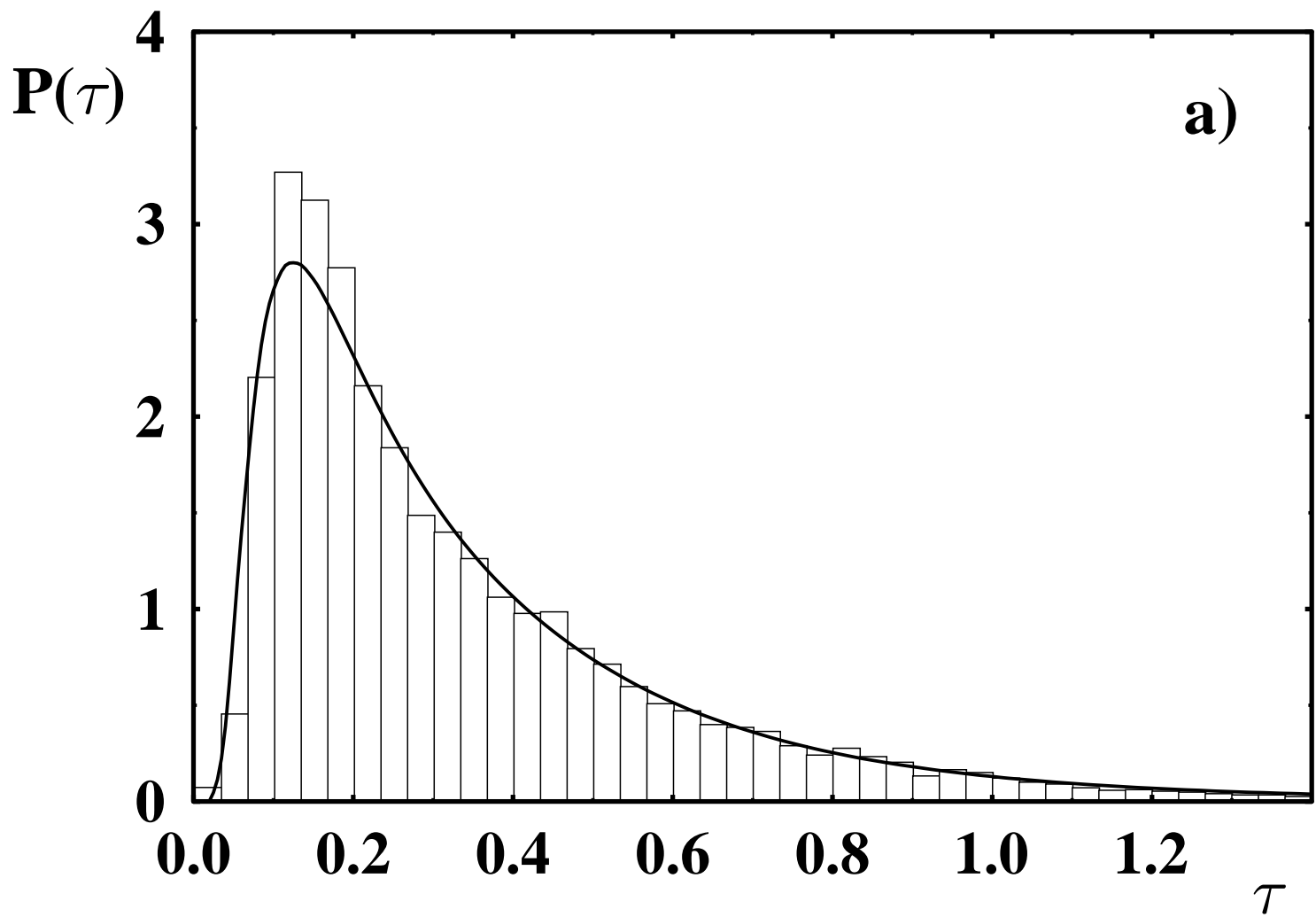
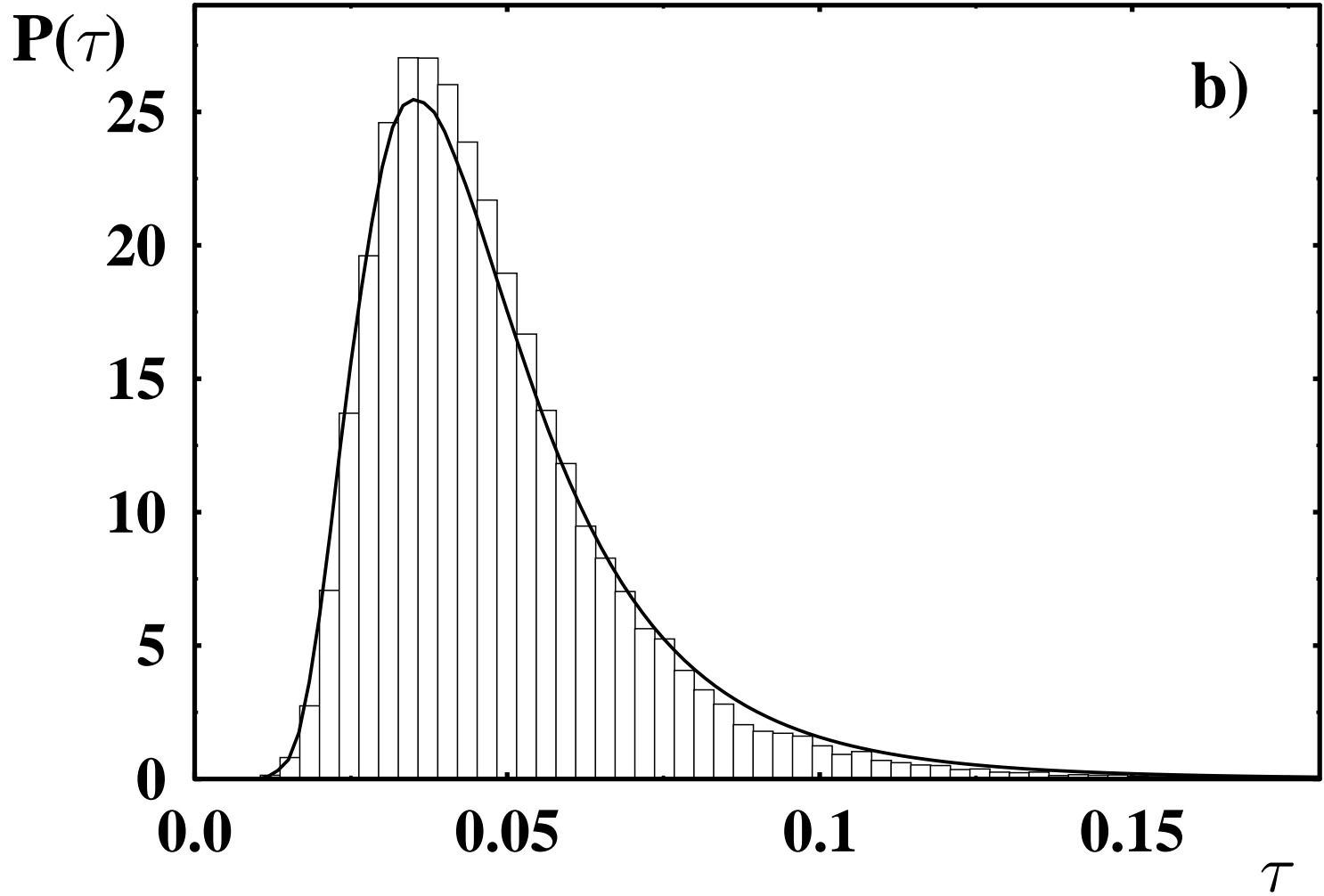
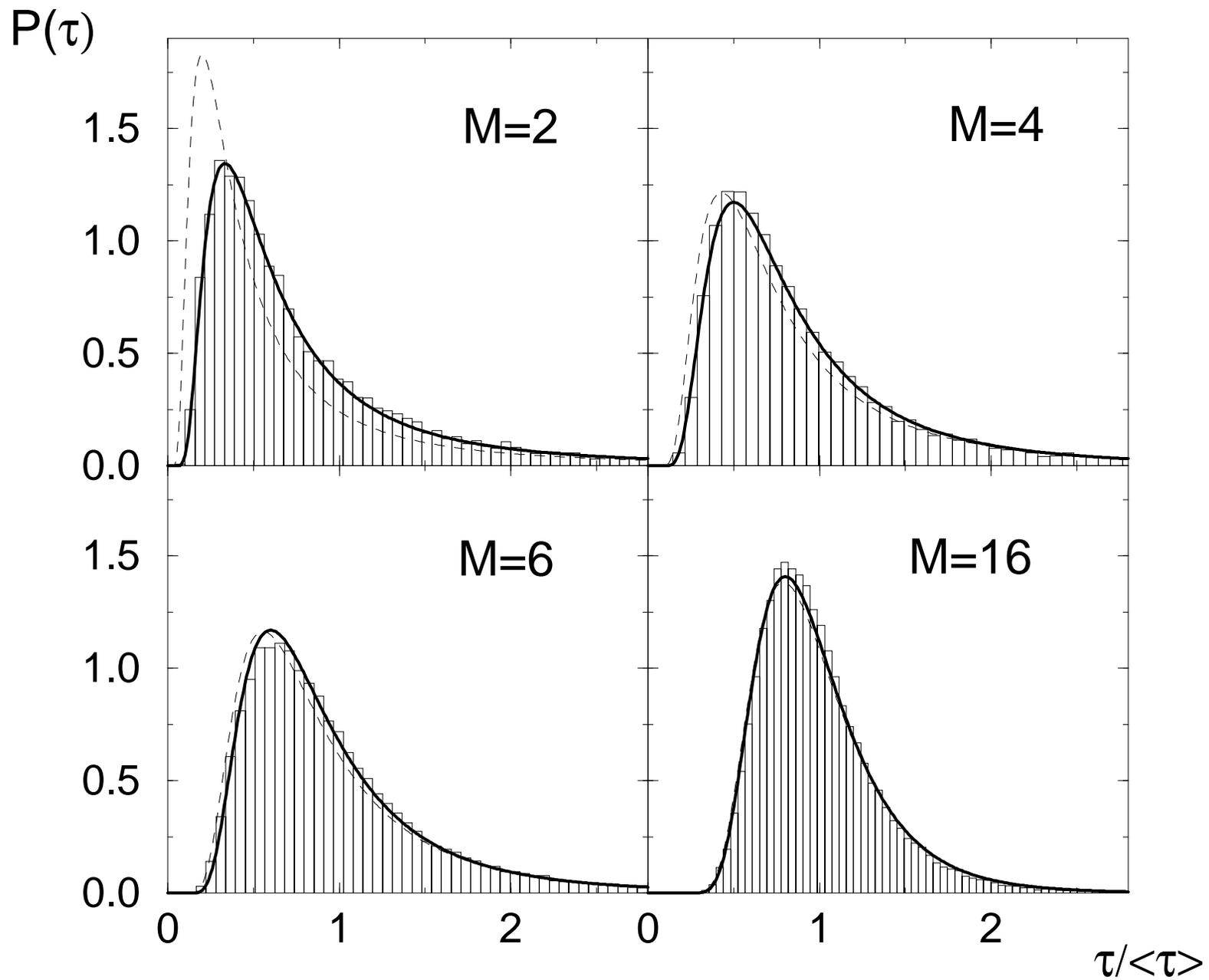
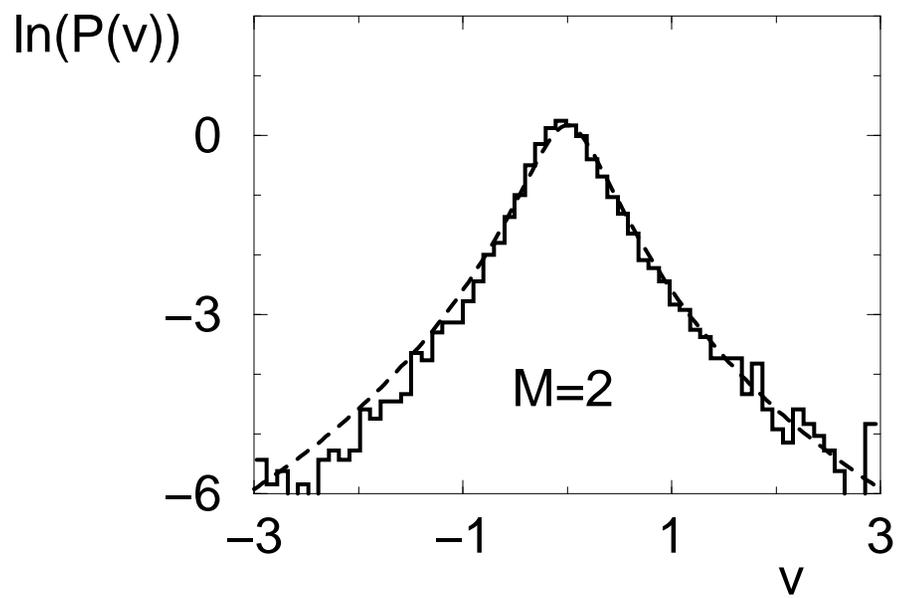
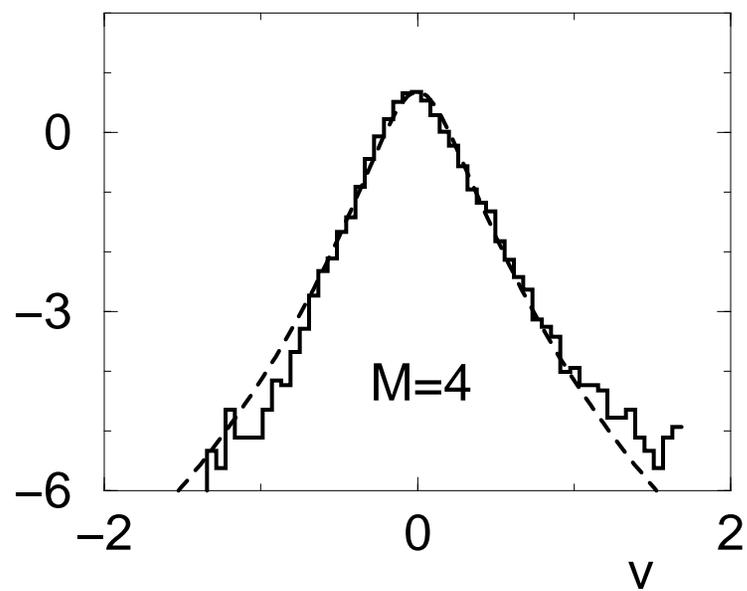
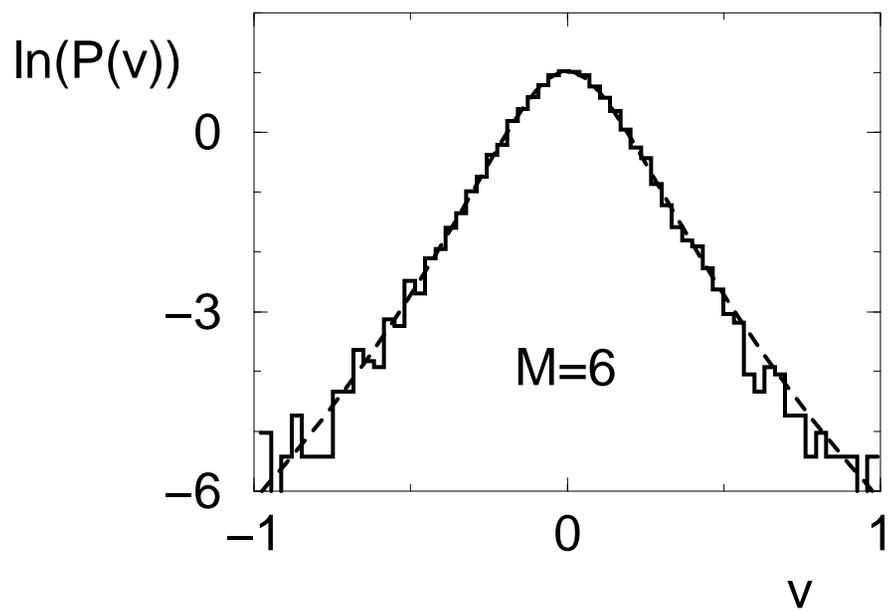
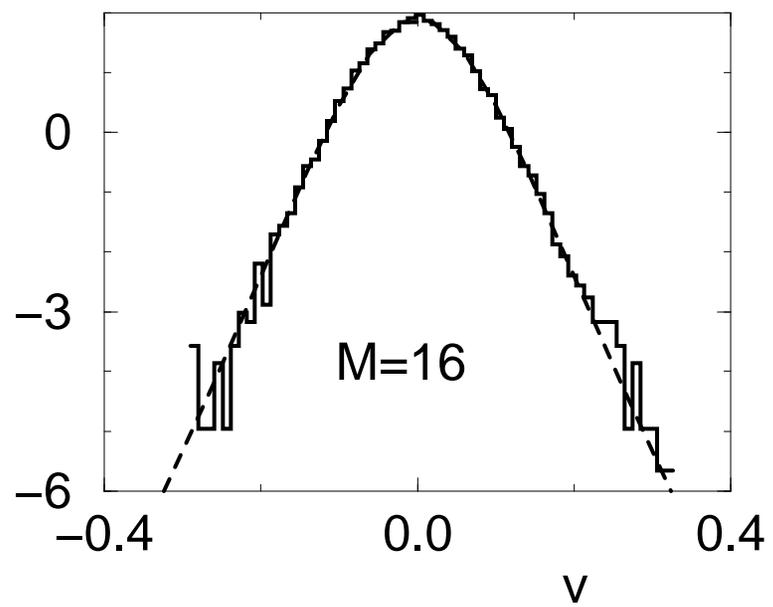
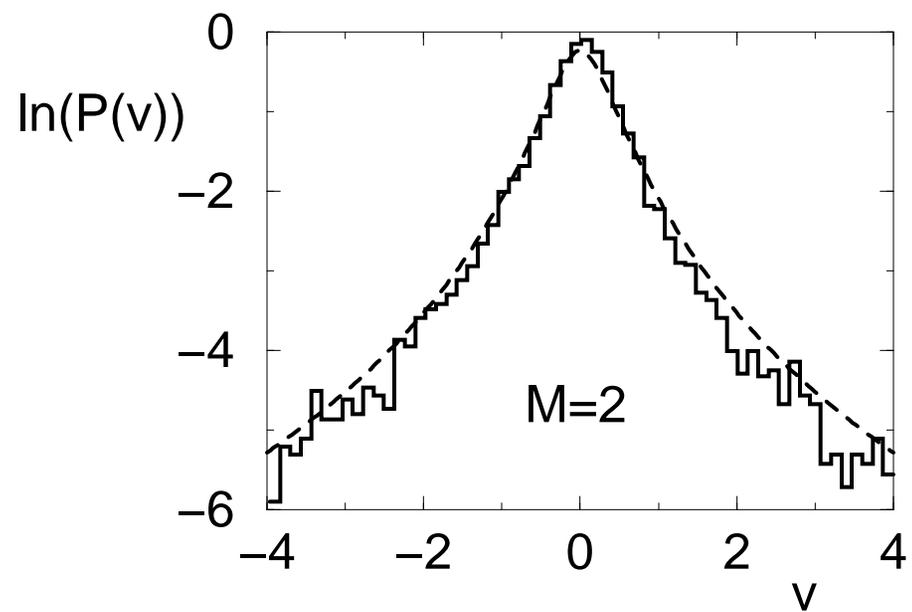
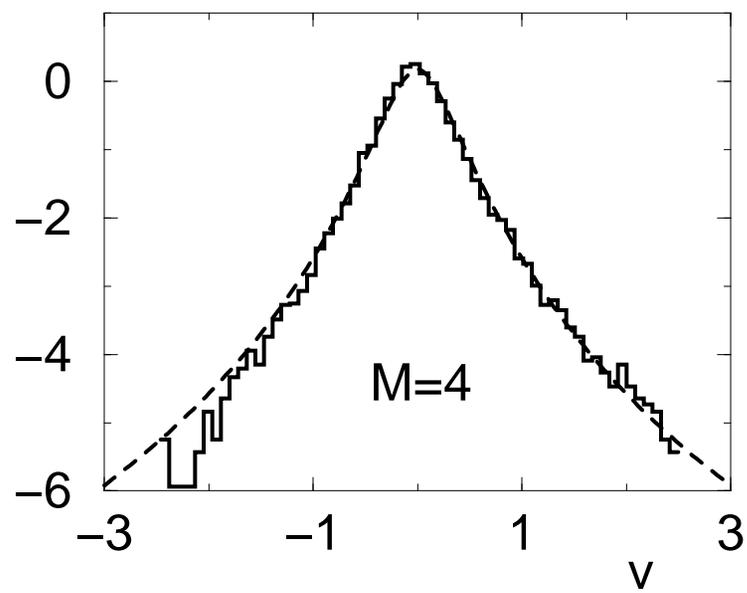
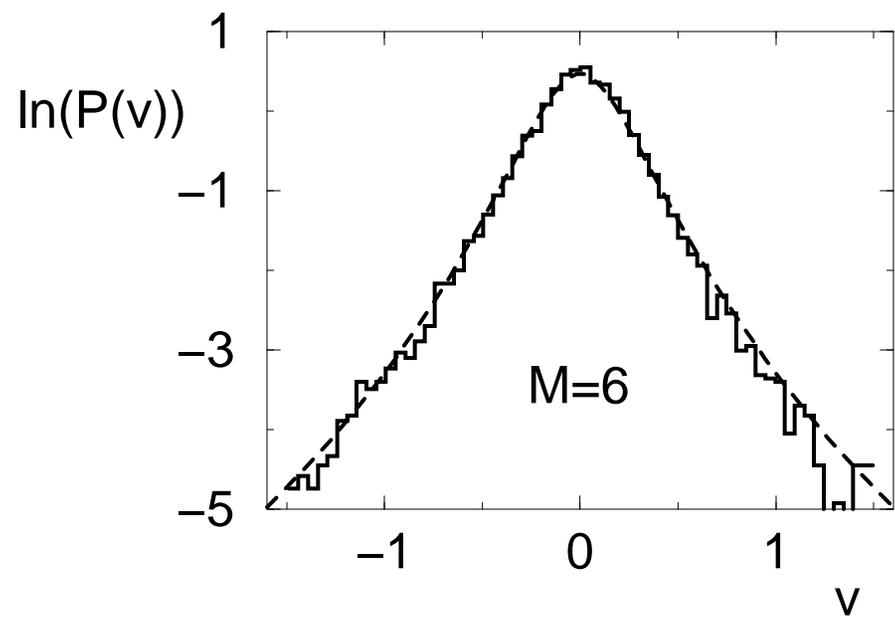
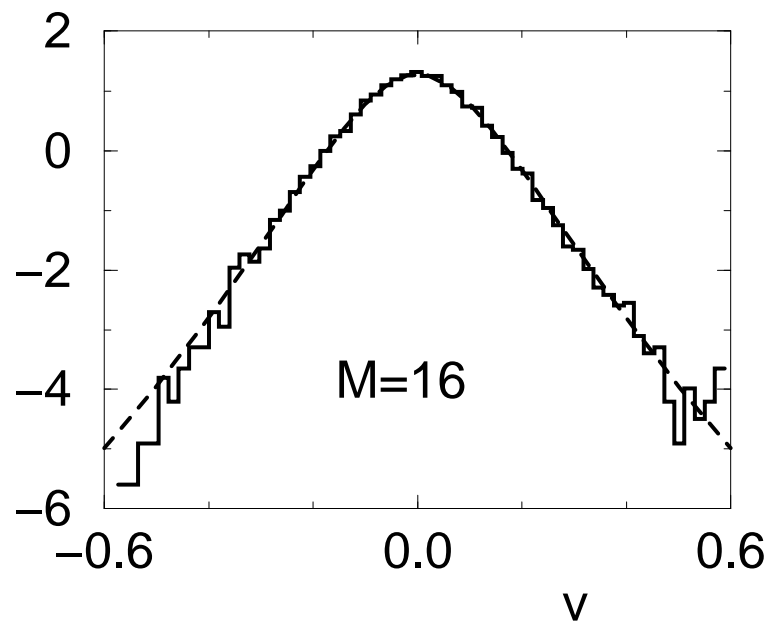
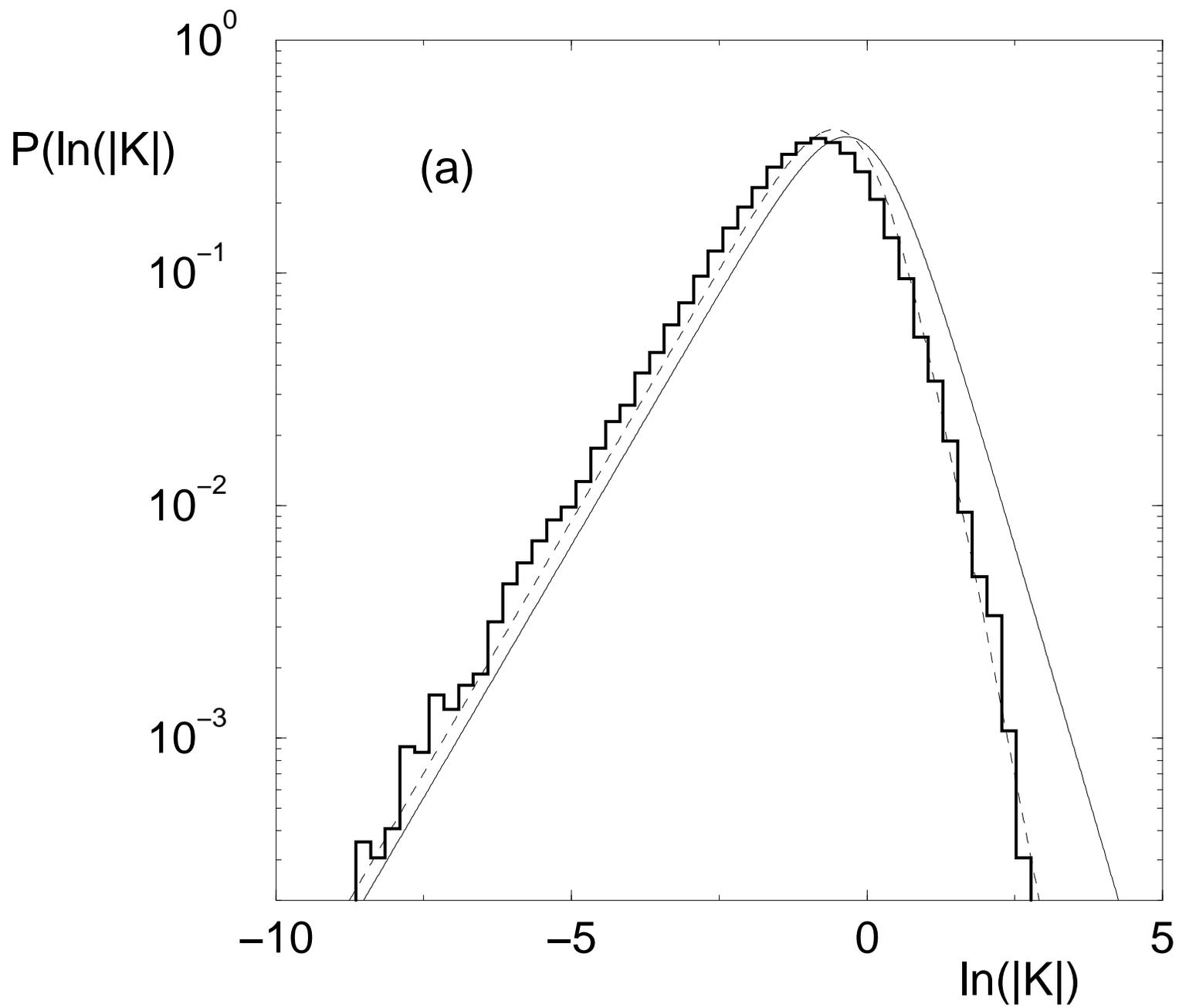
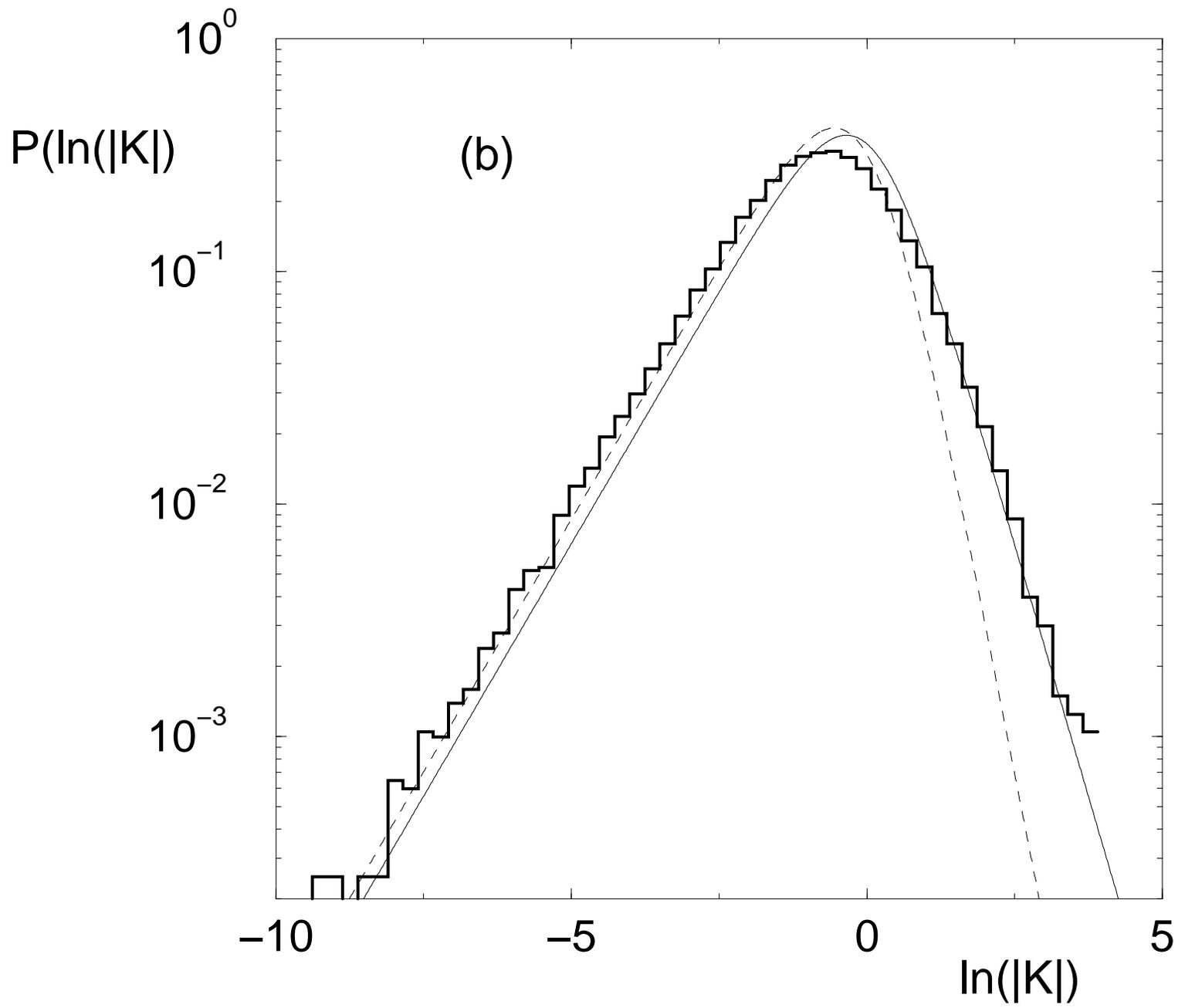
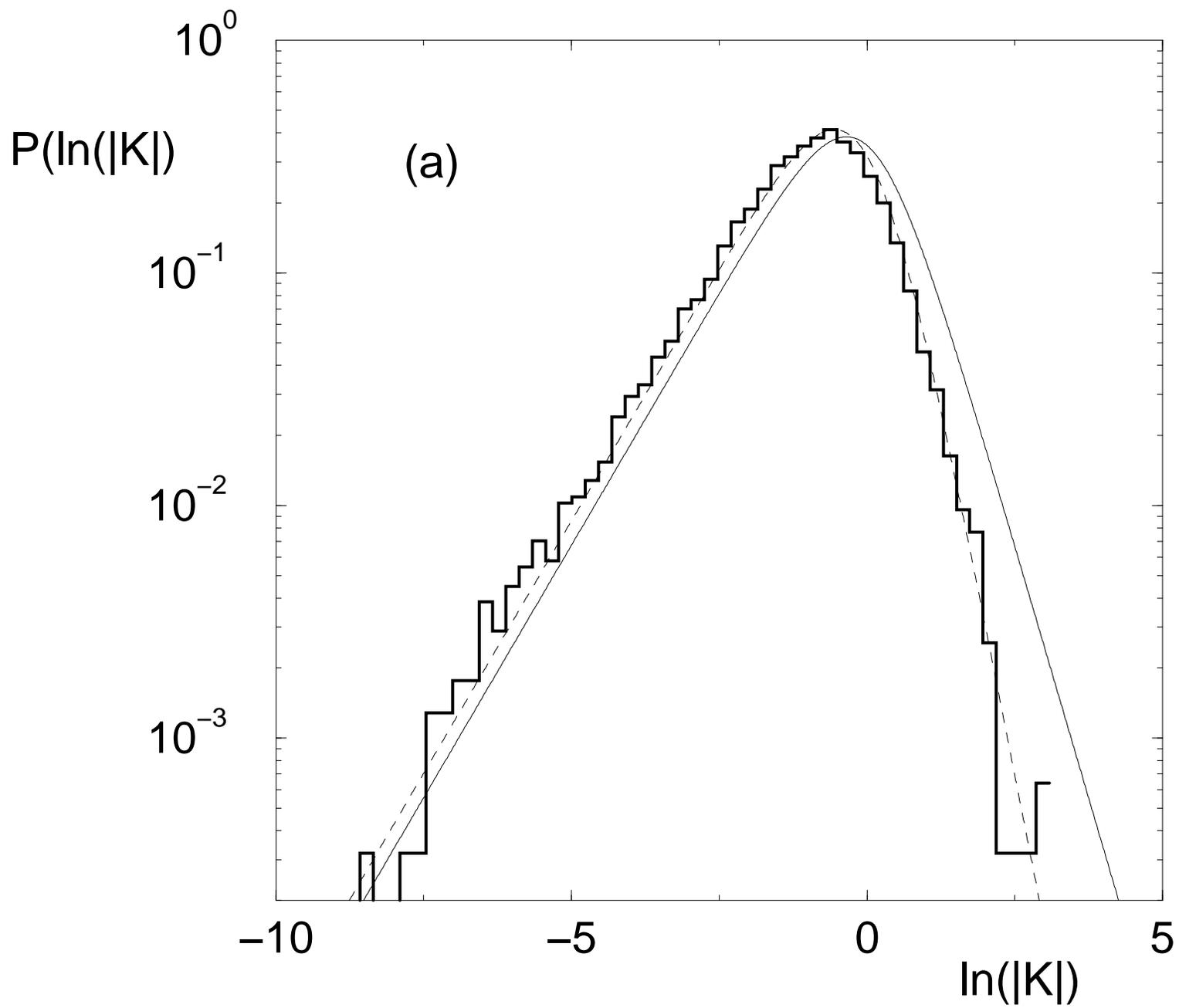
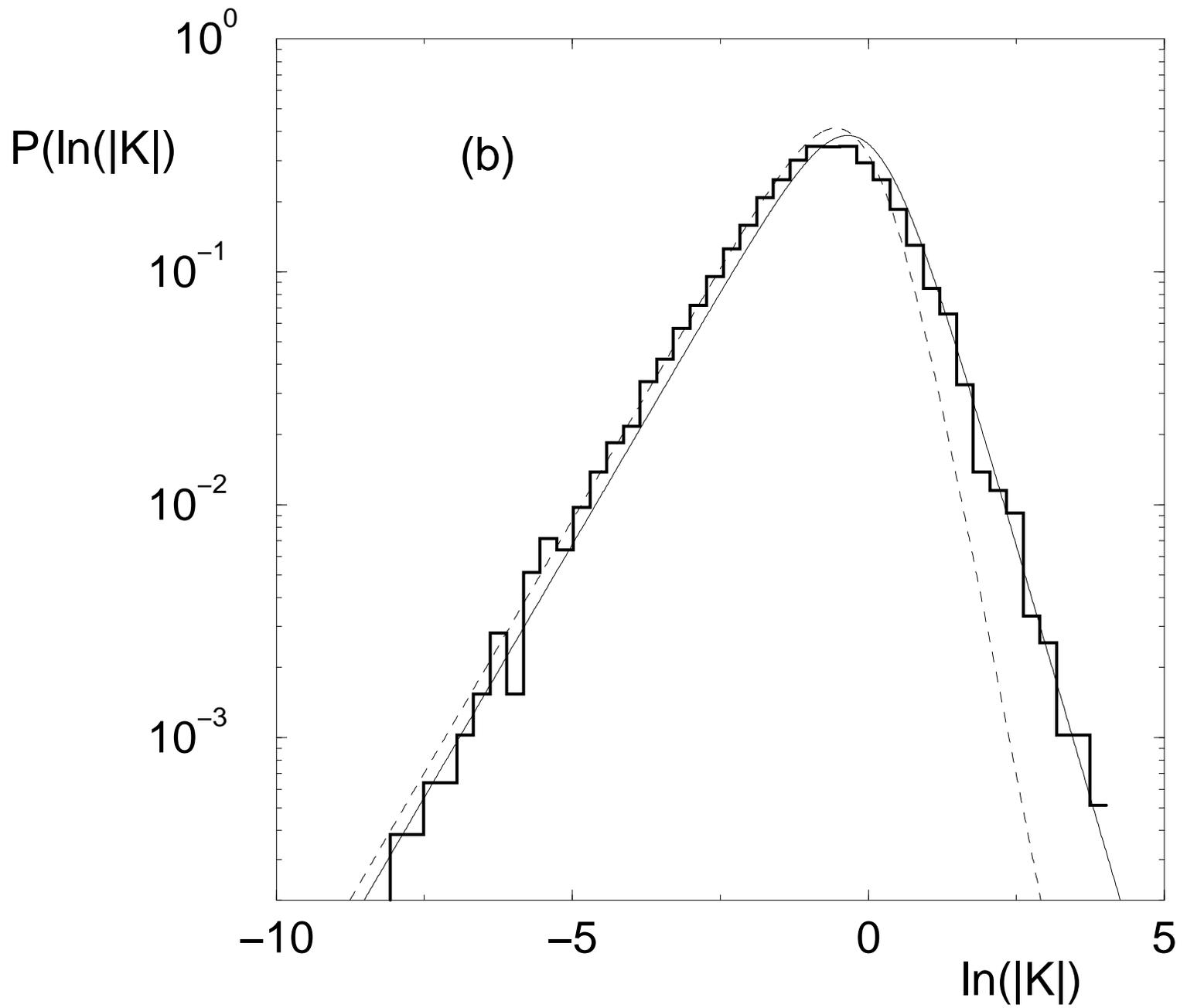
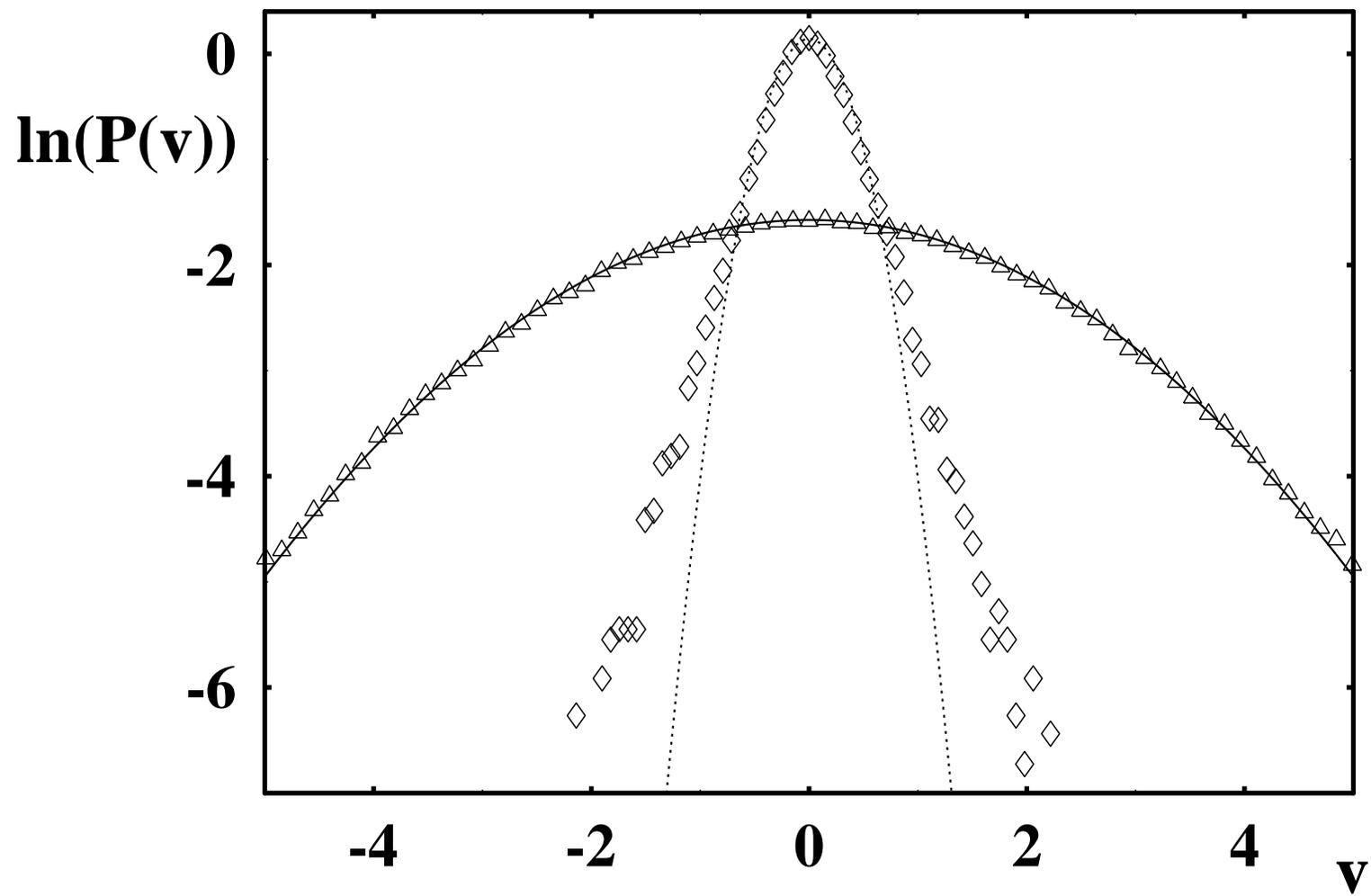
\end{document}